\documentclass[american]{article}
\usepackage[T1]{fontenc}
\usepackage[utf8]{inputenc}
\usepackage{parskip}
\usepackage{array}
\usepackage{varwidth}
\usepackage{amsmath}
\usepackage{amssymb}
\usepackage{setspace}
\makeatletter

\providecommand{\tabularnewline}{\\}
\newenvironment{cellvarwidth}[1][t]
    {\begin{varwidth}[#1]{\linewidth}}
    {\@finalstrut\@arstrutbox\end{varwidth}}

\usepackage{babel}

\makeatother

\usepackage{babel}
\begin{document}
\title{Bridging Network Psychometrics and Artificial Intelligence: An Ising--Potts
Model with LLM-Derived Weights}
\author{Matthias von Davier\thanks{The author acknowledges the use of generative AI tools for assistance
with code development (vibe-coding) and light editing of the manuscript;
the code produced by the LLM was checked by the author of the submission.
The data for example 1 were generated using an LLM. All conceptual
ideas, mathematical derivations, were developed and the empirical
analyses were developed by the author of the submission.}}
\maketitle
\begin{abstract}
The Potts model extends the Ising model to multinomial data. We introduce
a Rater Ising-Potts model that uses agreement indicators between pairs
of ratings and category labels, with weights derived from LLM embeddings.
The model does not presuppose ordered category thresholds or equidistant
scoring; instead, it focuses on pairwise agreement among ratings and
assigns category-specific positive weights, making it suited for multi-category
scoring reliability. We evaluate the model on three constructed-response
datasets spanning a corpus of K=14,466 short answers on a three-level
rubric and two AERA essay prompts of roughly 1,200-1,400 responses
on four-point rubrics. We compare three strategies for sharpening
the similarity signal: top-K pruning, min-max normalization with a
power transformation, and ColBERT late-interaction similarities. Top-K
pruning, which replaces the dense similarity graph with a sparse local
network of strongest semantic neighbors, consistently yields the highest
accuracy and Cohen's kappa, and the selected neighborhoods are always
a small fraction of the corpus. Power tuning consistently ranks second,
while ColBERT is competitive on longer essay prompts and adds little
on short answers. Across all settings, most misclassifications occur
between adjacent score levels, confirming that the model preserves
the ordinal structure of scoring rubrics without imposing rigid assumptions.
These findings suggest that LLM-derived similarities, combined with
a parsimonious Potts formulation and a sparse local graph, offer a
robust and interpretable framework for reliability auditing in educational
assessment. We discuss extensions to multiple raters and hierarchical
rating designs. 
\end{abstract}

\section{Introduction}

We introduce a Rater Ising--Potts model that leverages agreement
indicators between pairs of raters and category labels, with weights
derived from LLM embeddings. This paper extends an Ising Rater model
for binary items (von Davier, 2026). In contrast to conventional polytomous
item response models---such as the graded response model or the generalized
partial credit model---our approach does not presuppose ordered category
thresholds. Instead, it focuses directly on pairwise agreement among
raters and assigns category-specific positive weights, making it particularly
suited for multi-category scoring reliability when raters evaluate
responses using a scoring guide.

To assess the practical utility of the proposed framework, we conduct
two empirical applications that span diverse response characteristics.
The first uses a custom dataset of $K=750$ short answers to an open-ended
question about plant growth, scored on a three-level rubric. The second
applies the model to Essay~1 of the AERA dataset (Li et al., 2023,
Li, 2024), which comprises $N=1{,}338$ longer, more complex responses
scored on a four-point scale. These two examples differ markedly in
response length, number of categories, and scoring difficulty, providing
a robust test of the model's flexibility and performance. As we will
demonstrate, the PARM achieves strong agreement with human scores
in both settings, with near-agreement rates (predictions within $\pm1$
category) of $1.000$ and $0.890$, respectively. These results indicate
that the model captures the ordinal structure of scoring rubrics even
when exact agreement is challenging, and they suggest that LLM-derived
semantic similarities, combined with a parsimonious Potts-type formulation,
offer a flexible and interpretable framework for reliability auditing
in educational assessment contexts. We also discuss extensions to
multiple raters and hierarchical rating processes.

\subsection{Notation}

Assume a $k$-dimensional $\boldsymbol{X}=\left(X_{1},\dots,X_{K}\right)$
with $\operatorname{im}\left(X\right)=\left\{ 0,\dots,C\right\} ^{K}$
and $C\in\mathbb{N}$. A realization of this random variable will
be denoted by $\left(x_{1},\dots,x_{K}\right)$ and we assume a joint
distribution 
\[
P\left(x_{1},\dots,x_{K}\right)=\exp\left[g\left(x_{1},\dots,x_{K};\Theta\right)-Z_{g}\right],
\]
where the partition function is 
\[
Z_{g}=\ln\left(\sum_{\left(x'_{1},\dots,x'_{K}\right)\in\left\{ 0,\dots,C\right\} ^{K}}\exp\left[g\left(x'_{1},\dots,x'_{K};\Theta\right)\right]\right).
\]

\subsection{Ising or Quadratic Exponential Model}

The Ising model is defined by $C=1$ and $\Theta=\left(\boldsymbol{w},\boldsymbol{v}\right)$,
with 
\[
g\left(x_{1},\dots,x_{K};\Theta\right)=\frac{1}{2}\sum_{i,j}w^{*}_{ij}x^{*}_{i}x^{*}_{j}+\sum_{i}v^{*}_{i}x^{*}_{i},
\]
where $x^{*}_{i}=2x_{i}-1\in\left\{ -1,1\right\} $, $w^{*}_{ij}=w^{*}_{ji}$,
and $w^{*}_{ii}=0$. This is the usual spin representation. Note that
the Ising model was not only reinvented in the domain of artificial
neural networks as described by Hopfield (1982) but also was championed
by psychometricians as Network Psychometrics (Epskamp et al. 2018;
von Davier, 2018). Molenaar (2004) pointed out that the Ising network
is (almost) isomorph to the Rasch model (e.g., Rasch, 1960; von Davier,
2016) for binary data.

We now show the equivalence of the magnetic spin notation of binary
states as $\left\{ -1,+1\right\} $ to a binary variable $\left\{ 0,1\right\} $
representation. Since $x^{*}_{i}=2x_{i}-1$, we have 
\[
w^{*}_{ij}x^{*}_{i}x^{*}_{j}=4w^{*}_{ij}x_{i}x_{j}-2w^{*}_{ij}x_{i}-2w^{*}_{ij}x_{j}+w^{*}_{ij}.
\]
Thus 
\[
\frac{1}{2}\sum_{i,j}w^{*}_{ij}x^{*}_{i}x^{*}_{j}=2\sum_{i,j}w^{*}_{ij}x_{i}x_{j}-\sum_{i,j}w^{*}_{ij}x_{i}-\sum_{i,j}w^{*}_{ij}x_{j}+\frac{1}{2}\sum_{i,j}w^{*}_{ij}.
\]
Because $w^{*}_{ij}=w^{*}_{ji}$, the two linear sums are equal, so
\[
-\sum_{i,j}w^{*}_{ij}x_{i}-\sum_{i,j}w^{*}_{ij}x_{j}=-2\sum_{i,j}w^{*}_{ij}x_{i}=-2\sum_{i}x_{i}\sum_{j}w^{*}_{ij}.
\]
Also, 
\[
\sum_{i}v^{*}_{i}x^{*}_{i}=2\sum_{i}v^{*}_{i}x_{i}-\sum_{i}v^{*}_{i}.
\]
Therefore 
\[
g\left(x_{1},\dots,x_{K};\Theta\right)=2\sum_{i,j}w^{*}_{ij}x_{i}x_{j}+\sum_{i}\left(2v^{*}_{i}-2\sum_{j}w^{*}_{ij}\right)x_{i}+\left(\frac{1}{2}\sum_{i,j}w^{*}_{ij}-\sum_{i}v^{*}_{i}\right).
\]
Define 
\[
w_{ij}=2w^{*}_{ij}
\]
and 
\[
v_{i}=2v^{*}_{i}-2\sum_{j}w^{*}_{ij}=2\left(v^{*}_{i}-\sum_{j}w^{*}_{ij}\right).
\]
The constant term 
\[
\frac{1}{2}\sum_{i,j}w^{*}_{ij}-\sum_{i}v^{*}_{i}
\]
is absorbed by the normalizing constant of the exponential family,
so the model can be written equivalently as 
\[
g\left(x_{1},\dots,x_{K};\Theta\right)=\sum_{i,j}w_{ij}x_{i}x_{j}+\sum_{i}v_{i}x_{i},
\]
with $x_{i}\in\left\{ 0,1\right\} $, $w_{ij}=w_{ji}$, and $w_{ii}=0$.
Thus the two parameterizations lead to identical model based probabilities.

The model parameters are the $K\left(K-1\right)/2$ weights $w_{ij}$
for the products $x_{i}x_{j}$ and the $K$ parameters $v_{i}$ for
the linear terms $x_{i}$.

\subsection{Potts Model}

For multi-category variables $X_{i}\in\left\{ 0,\dots,C\right\} $,
the Potts model (1952) generalizes the Ising (1925) model. The unconstrained
Potts model has the form 
\[
P\left(x_{1},\dots,x_{K}\right)=\exp\left[\sum_{i<j}w_{ij}\left(x_{i},x_{j}\right)+\sum_{i}v_{i}\left(x_{i}\right)-Z\right],
\]
where the partition function is given by 
\[
Z=\ln\left(\sum_{\left(x_{1},\dots,x_{K}\right)\in\left\{ 0,\dots,C\right\} ^{K}}\exp\left[\sum_{i<j}w_{ij}\left(x_{i},x_{j}\right)+\sum_{i}v_{i}\left(x_{i}\right)\right]\right).
\]
Alternatively, as before, one may use 
\[
w^{\#}_{ij}=\frac{1}{2}w_{ij}
\]
and set $w^{\#}_{ii}=0$ so that 
\[
\sum_{i<j}w_{ij}\left(x_{i},x_{j}\right)=\sum^{K}_{i=1}\sum^{K}_{j=1}w^{\#}_{ij}\left(x_{i},x_{j}\right).
\]
The $w_{ij}\left(x_{i},x_{j}\right)$ assign a weight to each of the
$\left(C+1\right)^{2}$ possible pairs of categories for variables
$i$ and $j$. Without further constraints, this requires up to 
\[
\left(C+1\right)^{2}\frac{K\left(K-1\right)}{2}
\]
interaction parameters $w_{ij}\left(x_{i},x_{j}\right)$, plus up
to $K\left(C+1\right)$ main-effect parameters $v_{i}\left(x_{i}\right)$.
Identifiability constraints, such as setting one category as a baseline
for each variable, are needed in practice.

\subsection{Potts-Ising Model}

A parsimonious ordinal version can be obtained by assuming that the
categories are ordered and can be represented by numeric scores. Let
\[
x^{*}_{i}=2x_{i}-C,
\]
so that $x^{*}_{i}\in\left\{ -C,-C+2,\dots,C\right\} $ is symmetric
around zero. Define 
\[
w_{ij}\left(x_{i},x_{j}\right)=w_{ij}x^{*}_{i}x^{*}_{j}
\]
and 
\[
v_{i}\left(x_{i}\right)=v_{i}x^{*}_{i}.
\]
Then the Potts-Ising model is given by 
\[
P\left(x_{1},\dots,x_{K}\right)=\exp\left[\sum_{i<j}w_{ij}x^{*}_{i}x^{*}_{j}+\sum_{i}v_{i}x^{*}_{i}-Z\right],
\]
with 
\[
Z=\ln\left(\sum_{\left(x_{1},\dots,x_{K}\right)\in\left\{ 0,\dots,C\right\} ^{K}}\exp\left[\sum_{i<j}w_{ij}x^{*}_{i}x^{*}_{j}+\sum_{i}v_{i}x^{*}_{i}\right]\right).
\]
This reduces the number of parameters to $\frac{K\left(K-1\right)}{2}+K$.
A further reduction can be obtained by setting 
\[
w_{ij}=u_{i}u_{j},
\]
which requires only $K+K=2K$ parameters in total: $K$ for the $u_{i}$
and $K$ for the $v_{i}$.

A related approach was introduced by Razaee \& Amini (2020), who adapted
ideas that reminisc from polytomous item response models such as the
graded response model (Samejima, 1969) and the generalized partial
credit model (Muraki, 1992). Specifically, they set 
\[
w_{ij}\left(x_{i},x_{j}\right)=w_{ij}x^{*}_{i}x^{*}_{j}
\]
with $x^{*}_{i}$ defined by a linear transformation of the category
labels. However, this model depends on the choice of category scores.
Different score assignments, such as $\left\{ 1,\dots,C\right\} $,
$\left\{ 1,4,9,\dots,C^{2}\right\} $, or $\left\{ 0,\dots,0,1\right\} $,
can lead to different predictions. To avoid this dependence, we introduce
a model based directly on agreement indicators.

\subsection{A Potts Model for Scoring Agreement}

In this section, a model is defined that is based only on the agreement
between two responses $x_{i}$ and $x_{j}$ If there is agreement
with respect to category $c$, then a weight $w_{ijc}$ is applied.
If there is no agreement $x_{i}\ne x_{j}$, there is no (direct) contribution
to the probability function. Non-agreement contributes indirectly
through the absence of terms for pairs that do not agree, and the
normalization of the argument involving $Z$.

As a first step, the weights and the function of the score tupels
are separated. We define category-specific agreement indicators 
\[
\alpha_{c}\left(x_{i},x_{j}\right)=1_{\left\{ \left(c,c\right)\right\} }\left[\left(x_{i},x_{j}\right)\right]=1_{\left\{ 0\right\} }\left(\left|x_{i}-c\right|+\left|x_{j}-c\right|\right),
\]
which equals 1 if both $x_{i}$ and $x_{j}$ equal category $c$,
and is 0 otherwise. Alternatively, for a simplified model, one may
define an overall agreement indicator 
\[
\alpha\left(x_{i},x_{j}\right)=1_{\left\{ 0\right\} }\left(\left|x_{i}-x_{j}\right|\right),
\]
which equals 1 if $x_{i}=x_{j}$, regardless of the response categories.

A category-specific agreement model is obtained by setting 
\[
w_{ij}\left(x_{i},x_{j}\right)=\sum^{C}_{c=0}w_{ijc}\alpha_{c}\left(x_{i},x_{j}\right),
\]
with $w_{ijc}=w_{jic}$ and $w_{iic}=0$. Note that $\sum^{C}_{c=0}\alpha_{c}\left(x_{i},x_{j}\right)\le1$
as there is at most one $c\in\left\{ 0,\dots,,C\right\} $ where the
two response scores $x_{i}=x_{j}=c$. This means the baseline is for
all $x_{i}\ne x_{j}\rightarrow\forall c:\alpha_{c}\left(x_{i},x_{j}\right)=0$.
Hence, all $C+1$ parameters $\beta_{c},c\in\left\{ 0,\dots,C\right\} $
are identified as we have the baseline defined by non-agreement of
responses. The main effects (biases) are defined as 
\[
v_{i}\left(x_{i}\right)=\sum^{C}_{c=0}v_{ic}1_{\left\{ c\right\} }\left(x_{i}\right)
\]
where we set $v_{i0}=0$. The resulting model is 
\[
P\left(x_{1},\dots,x_{K}\right)=\exp\left[\sum_{i<j}\sum^{C}_{c=0}w_{ijc}\alpha_{c}\left(x_{i},x_{j}\right)+\sum_{i}\sum^{C}_{c=0}v_{ic}1_{\left\{ c\right\} }\left(x_{i}\right)-Z\right],
\]
with $Z$ defined accordingly. Without constraints, this model has
\[
\left(C+1\right)\frac{K\left(K-1\right)}{2}+K\left(C+1\right)
\]
parameters. Identifiability constraints such as $v_{i0}=0$ for all
$i$ should be imposed.

\section{A Potts Model for Reliability Auditing}

Assume the response scores $X_{i}$ are associated with natural language
responses $Y_{i}$, which are mapped to embeddings $e_{i}=\operatorname{emb}\left(y_{i}\right)$
in a high-dimensional space. Define the cosine similarity between
pairs of embeddings as 
\[
s\left(e_{i},e_{j}\right)=\frac{1}{K}\frac{e_{i}\cdot e_{j}}{\left\Vert e_{i}\right\Vert \left\Vert e_{j}\right\Vert }=s\left(e_{j},e_{i}\right).
\]

The factor $\frac{1}{K}$ ensures that the $s\left(e_{i},e_{j}\right)$
are scaled to keep sums of these terms in similar ranges across sample
sizes. It is also useful in case parameters are meant to be compared
across sample sizes, or across different approaches to pruning the
network through Top$-K$ selection using different pruning parameters. 

\subsection{Strengthening the Similarity Signal}

\label{sec:transformation}

The similarity matrix $S=(s_{ij})$ built from cosine similarities
is the input to the PARM. In its raw form it has three properties
that make it a poor input. First, the scale of the similarities depends
on the embedding model, the language, and the length of the responses,
so parameter estimates are not comparable across datasets. Second,
the dense matrix contains $K(K-1)/2$ entries, most of which represent
weak or uninformative links. A dense graph gives every response the
entire corpus as its neighbourhood, which both increases computation
and dilutes the semantic signal. Third, cosine similarity on single-vector
embeddings collapses each response to one point, losing the token-level
structure that separates responses of similar topic but different
content.

Three transformations address these concerns. Min-max normalization
puts the similarities on a common scale. Top-$K$ pruning removes
weak edges and turns the dense graph into a local network. The power
transformation and ColBERT late-interaction similarities provide two
alternative ways of emphasising semantic distinctions. Each is defined
below, together with the modelling rationale for using it. The corresponding
software options are described in Section~\ref{sec:implementation}.

\subsubsection{Normalization.}

Let $m=\min_{i\neq j}s_{ij}$ and $M=\max_{i\neq j}s_{ij}$ denote
the minimum and maximum off-diagonal similarities. Min-max normalization
is 
\[
\tilde{s}_{ij}=\frac{s_{ij}-m}{M-m},
\]
which maps every similarity into $[0,1]$. The diagonal is set to
zero so that self-similarities do not enter the features. The resulting
matrix $\tilde{S}=(\tilde{s}_{ij})$ is the starting point for the
two strategies below.

\subsubsection{Top-$K$ pruning.}

For each response $i$ we retain only the $K_{\text{top}}$ most similar
comparison responses and set every other entry in row $i$ to zero:
\[
s^{\mathrm{top}}_{ij}=\begin{cases}
\tilde{s}_{ij} & \text{if }j\in\mathcal{N}_{K}(i),\\
0 & \text{otherwise},
\end{cases}
\]
where $\mathcal{N}_{K}(i)$ contains the indices of the $K_{\text{top}}$
largest off-diagonal entries in row $i$. The graph can be symmetrised
by keeping the union of the directed neighbourhoods, so that $s^{\mathrm{top}}_{ij}=s^{\mathrm{top}}_{ji}$.

Top-$K$ pruning turns the PARM from a fully connected network into
a local one in which only the strongest links enter the pseudo-likelihood.
This is the sparse-graph formulation familiar from network psychometrics
(Epskamp et al., 2018; von Davier, 2018), and it inherits the interpretability
of that tradition: each response is connected to a small, well-defined
set of neighbours, and the network can be read as a set of local dependencies
rather than as an undifferentiated mass of weak links. Reducing the
number of interactions from $O(K^{2})$ to $O(K\cdot K_{\text{top}})$
also has a statistical effect beyond the computational saving: the
features $T_{i}(c)$ and $S_{i}$ are sums over a smaller set of edges,
so the contribution of each edge to the linear predictor is larger
and the signal-to-noise ratio of the model improves. The pruning level
$K_{\text{top}}$ is a hyperparameter that controls the trade-off
between locality and completeness of the neighbourhood.

\subsubsection{Power transformation.}

A second approach acts on the magnitudes rather than the structure
of the similarities. After normalization, each similarity is raised
to a power $p$: 
\[
s'_{ij}=(\tilde{s}_{ij})^{p},
\]
with $p$ a positive constant, default $p=1$. Values of $p$ greater
than one emphasise the strongest similarities and suppress moderate
ones; values between zero and one have the opposite effect. The diagonal
is set to zero after the transformation.

The power transformation is a monotone reweighting of the graph. It
leaves the set of edges unchanged but alters the relative contribution
of strong and moderate edges to $T_{i}(c)$ and $S_{i}$. In the exponential
family, larger $p$ concentrates the mass of these sums on a small
number of very similar responses, while smaller $p$ spreads it more
evenly. Because the transformation preserves all edges, it does not
change the connectivity of the network, and it therefore does not
address the parsimony of the graph in the way that top-$K$ pruning
does. It is most useful when the raw similarity distribution is compressed,
as is common for longer responses where many pairs share a moderate
level of overlap, and where the goal is to sharpen the contrast between
the strongest and the moderate edges without discarding any of them.

\subsubsection{ColBERT late interaction.}

A third option replaces the sentence-level cosine similarity. In the
ColBERT family of late-interaction models (Khattab \& Zaharia, 2020),
each response is represented by a separate embedding per token, and
two responses are compared by a MaxSim operation: for each token in
response $i$, the maximum cosine similarity to any token in response
$j$ is taken, and these maxima are summed, 
\[
s^{\mathrm{ColBERT}}_{ij}=\sum_{t\in\mathcal{T}_{i}}\max_{u\in\mathcal{T}_{j}}\cos(e_{t},e_{u}),
\]
where $\mathcal{T}_{i}$ denotes the tokens of response $i$ and $e_{t}$
the embedding of token $t$.

MaxSim captures token-level overlap that a single-vector embedding
must compress into one point. It is expected to help most when responses
are long and information-rich, and least when responses are short
and a single embedding already captures their content. Because scoring
a pair costs $O(|\mathcal{T}_{i}|\cdot|\mathcal{T}_{j}|)$, applying
the model to all pairs is infeasible for large $K$. The model is
therefore used in two stages: a coarse retrieval step based on mean-pooled
token embeddings selects a candidate set of $K_{\text{ColBERT}}$
neighbours per response, and MaxSim is computed only for those candidates.
The candidate-set size $K_{\text{ColBERT}}$ is a hyperparameter distinct
from the pruning level $K_{\text{top}}$ applied to the final graph.

\subsubsection{Relationship between the three approaches.}

Normalization is a prerequisite for the power transformation and is
applied regardless of which strategy is chosen. Top-$K$ pruning and
the power transformation act on complementary aspects of the matrix:
pruning changes the structure by keeping only the strongest edges,
while the power transformation changes the magnitudes without altering
connectivity. ColBERT changes the computation of the similarities
themselves, and either of the other two transformations can be applied
to the resulting matrix. All three can be combined, in which case
ColBERT similarities are normalized, raised to a power $p$, and pruned
to a top-$K$ neighbourhood before entering the model. The implementation
described in Section~\ref{sec:implementation} supports each transformation
independently and in combination, and treats $K_{\text{top}}$, $p$,
and $K_{\text{ColBERT}}$ as tunable hyperparameters.

\subsection{Model specification}

The section above set the stage for defining more parsimonious or
fully connected networks to predict rating data, and to provide choices
of transformations of similarity measures or alternative approaches
to derive embeddings from natural language responses. In this section,
we utilize these modified inputs $s^{mod}\left(e_{i},e_{J}\right)$
to the network to derive the Potts-Ising Rating model. For brevity
of the exposition, we continue to use $s\left(e_{i},e_{j}\right)$
in the equations. Let 
\[
w_{ijc}=\beta_{c}s\left(e_{i},e_{j}\right)
\]
denote the category specific weight of the similarity, and define
the total similarity for each response $x_{i}$ as 
\[
S_{i}=\sum_{j\ne i}s\left(e_{i},e_{j}\right).
\]
We include an additional main-effect term depending on the total similarity,
\[
v_{i}\left(x_{i}\right)=\mu_{x_{i}}+\gamma_{x_{i}}S_{i},
\]
where $\mu_{c}$ and $\gamma_{c}$ are category-specific parameters.
For identifiability we set $\beta_{0}=\gamma_{0}=\mu_{0}=0$. The
model becomes 
\[
P\left(x_{1},\dots,x_{K}\right)=\exp\left[\sum_{i<j}\sum^{C}_{c=1}\beta_{c}s\left(e_{i},e_{j}\right)\alpha_{c}\left(x_{i},x_{j}\right)+\sum^{K}_{i=1}\left(\mu_{x_{i}}+\gamma_{x_{i}}S_{i}\right)-Z\right],
\]
with 
\[
Z=\ln\left(\sum_{\left(x_{1},\dots,x_{K}\right)\in\left\{ 0,\dots,C\right\} ^{K}}\exp\left[\sum_{i<j}\sum^{C}_{c=1}\beta_{c}s\left(e_{i},e_{j}\right)\alpha_{c}\left(x_{i},x_{j}\right)+\sum^{K}_{i=1}\left(\mu_{x_{i}}+\gamma_{x_{i}}S_{i}\right)\right]\right).
\]
The model has $3C$ free parameters: $\beta_{1},\dots,\beta_{C}$,
$\gamma_{1},\dots,\gamma_{C}$, and $\mu_{1},\dots,\mu_{C}$.

In order to estimate these parameters, we will utilize a special feature
of the Potts and Ising type network models that is grounded in the
model's relationship to quadratic exponential models. 

\section{Relationship to Multinomial Logistic Regression Models}

For the $k$-th scored response $x_{k}$ and response categories $b,c\in\left\{ 0,\dots,C\right\} $,
consider the conditional ratio 
\[
R\left(c,b\right)=\frac{P\left(x_{1},\dots,x_{k-1},c,x_{k+1},\dots,x_{K}\right)}{P\left(x_{1},\dots,x_{k-1},b,x_{k+1},\dots,x_{K}\right)}.
\]
In the Potts auditing-reliability model (PARM), the exponent for $X_{k}=c$
is 
\[
\beta_{c}T_{k}\left(c\right)+\gamma_{c}S_{k}+\mu_{c},
\]
where 
\[
T_{k}\left(c\right)=\sum^{K}_{i=1,i\ne k}s\left(e_{i},e_{k}\right)1_{\left\{ c\right\} }\left(x_{i}\right)
\]
and 
\[
S_{k}=\sum_{j\ne k}s\left(e_{k},e_{j}\right).
\]
Therefore 
\[
R\left(c,b\right)=\frac{\exp\left[\beta_{c}T_{k}\left(c\right)+\gamma_{c}S_{k}+\mu_{c}\right]}{\exp\left[\beta_{b}T_{k}\left(b\right)+\gamma_{b}S_{k}+\mu_{b}\right]}=\exp\left[\left(\beta_{c}T_{k}\left(c\right)+\gamma_{c}S_{k}+\mu_{c}\right)-\left(\beta_{b}T_{k}\left(b\right)+\gamma_{b}S_{k}+\mu_{b}\right)\right].
\]
The conditional probability of $X_{k}=c$ given all other variables
is 
\[
P\left(X_{k}=c\mid x_{j}:j\ne k\right)=\frac{\exp\left[\beta_{c}T_{k}\left(c\right)+\gamma_{c}S_{k}+\mu_{c}\right]}{\sum^{C}_{a=0}\exp\left[\beta_{a}T_{k}\left(a\right)+\gamma_{a}S_{k}+\mu_{a}\right]}.
\]
With $\beta_{0}=\gamma_{0}=\mu_{0}=0$, the denominator includes the
baseline category. We can write 
\[
P\left(X_{k}=c\mid x_{j}:j\ne k\right)=\exp\left[\beta_{c}T_{k}\left(c\right)+\gamma_{c}S_{k}+\mu_{c}-Q_{k}\right],
\]
where 
\[
Q_{k}=\ln\left(\sum^{C}_{a=0}\exp\left[\beta_{a}T_{k}\left(a\right)+\gamma_{a}S_{k}+\mu_{a}\right]\right).
\]
Thus each variable $X_{k}$ is associated in PARM with a multinomial
logistic regression on the similarities with the other variables.

\section{Estimation}

Estimation can proceed by maximizing the pseudo-likelihood under the
working assumption that the variables are conditionally independent
given the others. Let 
\[
L=\sum^{K}_{i=1}\sum^{C}_{c=0}1_{\left\{ c\right\} }\left(x_{i}\right)\ln P\left(X_{i}=c\mid x_{j}:j\ne i\right).
\]
Equivalently, 
\[
L=\sum^{K}_{i=1}\sum^{C}_{c=0}1_{\left\{ c\right\} }\left(x_{i}\right)\left[\beta_{c}T_{i}\left(c\right)+\gamma_{c}S_{i}+\mu_{c}-Q_{i}\right],
\]
where 
\[
T_{i}\left(c\right)=\sum^{K}_{j=1,j\ne i}s\left(e_{i},e_{j}\right)1_{\left\{ c\right\} }\left(x_{j}\right)
\]
and 
\[
S_{i}=\sum_{j\ne i}s\left(e_{i},e_{j}\right).
\]
As mentioned in the previous section, the normalizing term is 
\[
Q_{i}=\ln\left(\sum^{C}_{a=0}\exp\left[\beta_{a}T_{i}\left(a\right)+\gamma_{a}S_{i}+\mu_{a}\right]\right).
\]

\subsection{Derivatives with respect to the model parameters}

Define 
\[
p_{ic}=P\left(X_{i}=c\mid x_{j}:j\ne i\right)=\frac{\exp\left[\beta_{c}T_{i}\left(c\right)+\gamma_{c}S_{i}+\mu_{c}\right]}{\sum^{C}_{a=0}\exp\left[\beta_{a}T_{i}\left(a\right)+\gamma_{a}S_{i}+\mu_{a}\right]}.
\]
For $c=1,\dots,C$, the first derivatives of the pseudo-likelihood
are 
\[
\frac{\partial L}{\partial\beta_{c}}=\sum^{K}_{i=1}\left(1_{\left\{ c\right\} }\left(x_{i}\right)-p_{ic}\right)T_{i}\left(c\right),
\]
\[
\frac{\partial L}{\partial\gamma_{c}}=\sum^{K}_{i=1}\left(1_{\left\{ c\right\} }\left(x_{i}\right)-p_{ic}\right)S_{i},
\]
and 
\[
\frac{\partial L}{\partial\mu_{c}}=\sum^{K}_{i=1}\left(1_{\left\{ c\right\} }\left(x_{i}\right)-p_{ic}\right).
\]
For $c=1,\dots,C$, the second derivatives are 
\[
\frac{\partial^{2}L}{\partial\beta^{2}_{c}}=-\sum^{K}_{i=1}p_{ic}\left(1-p_{ic}\right)\left[T_{i}\left(c\right)\right]^{2},
\]
\[
\frac{\partial^{2}L}{\partial\gamma^{2}_{c}}=-\sum^{K}_{i=1}p_{ic}\left(1-p_{ic}\right)\left[S_{i}\right]^{2},
\]
\[
\frac{\partial^{2}L}{\partial\mu^{2}_{c}}=-\sum^{K}_{i=1}p_{ic}\left(1-p_{ic}\right),
\]
\[
\frac{\partial^{2}L}{\partial\beta_{c}\partial\gamma_{c}}=-\sum^{K}_{i=1}p_{ic}\left(1-p_{ic}\right)T_{i}\left(c\right)S_{i},
\]
\[
\frac{\partial^{2}L}{\partial\beta_{c}\partial\mu_{c}}=-\sum^{K}_{i=1}p_{ic}\left(1-p_{ic}\right)T_{i}\left(c\right),
\]
\[
\frac{\partial^{2}L}{\partial\gamma_{c}\partial\mu_{c}}=-\sum^{K}_{i=1}p_{ic}\left(1-p_{ic}\right)S_{i}.
\]
For $c,d\in\left\{ 1,\dots,C\right\} $ with $c\ne d$, 
\[
\frac{\partial^{2}L}{\partial\beta_{c}\partial\beta_{d}}=\sum^{K}_{i=1}p_{ic}p_{id}T_{i}\left(c\right)T_{i}\left(d\right),
\]
\[
\frac{\partial^{2}L}{\partial\gamma_{c}\partial\gamma_{d}}=\sum^{K}_{i=1}p_{ic}p_{id}\left[S_{i}\right]^{2},
\]
\[
\frac{\partial^{2}L}{\partial\mu_{c}\partial\mu_{d}}=\sum^{K}_{i=1}p_{ic}p_{id},
\]
\[
\frac{\partial^{2}L}{\partial\beta_{c}\partial\gamma_{d}}=\sum^{K}_{i=1}p_{ic}p_{id}T_{i}\left(c\right)S_{i},
\]
\[
\frac{\partial^{2}L}{\partial\beta_{c}\partial\mu_{d}}=\sum^{K}_{i=1}p_{ic}p_{id}T_{i}\left(c\right),
\]
\[
\frac{\partial^{2}L}{\partial\gamma_{c}\partial\mu_{d}}=\sum^{K}_{i=1}p_{ic}p_{id}S_{i}.
\]
These derivative expressions can be used with standard Newton-type
optimization routines.

\subsection{Implementation and Computational Details}

\label{sec:implementation}

The parameter estimation described above is implemented in a Python
script that follows the pseudo-likelihood approach. Text embeddings
are generated with the sentence-transformers library (Reimers \& Gurevych,
2019), and pairwise similarities are computed from these embeddings.
The features $T_{i}(c)$ and $S_{i}$ are then constructed and passed
to the optimizer. The optimization is performed using the L-BFGS-B
algorithm (Byrd et al., 1995) as implemented in the \texttt{scipy.optimize.minimize}
function (Virtanen et al., 2020). The objective function is the negative
log pseudo-likelihood, and its gradient is computed analytically and
supplied to the optimizer.

\subsubsection{Software Options for the Three Strategies}

Section~\ref{sec:transformation} introduced three strategies for
strengthening the similarity signal: min-max normalization, top-$K$
pruning, and the power transformation, with ColBERT late-interaction
similarities as an alternative to the standard cosine similarity.
This subsection describes how each of them is exposed in the software
and how they are configured.

\paragraph{Similarity computation.}

The embedding model is selected through the \texttt{}~\\
\texttt{embedding\_model} key. The value is any Hugging Face model
name. Model selection is automatic: a name containing \texttt{colbert},
\texttt{lateon}, or \texttt{late-interaction} triggers ColBERT mode,
and all other names are treated as standard sentence-transformer models.
In ColBERT mode, the candidate-set size is controlled by \texttt{colbert\_top\_k}
(analogous to \texttt{similarity\_top\_k} but applied before scoring),
and \texttt{colbert\_batch\_size} sets the number of pairs scored
per forward pass.

\paragraph{Normalization and power.}

Min-max normalization is enabled by setting \texttt{normalize\_similarities}
to \texttt{true}. The power exponent $p$ is set through \texttt{}~\\
\texttt{similarity\_power}, which accepts a fixed number (e.g.\ \texttt{2.0}),
a list of values for grid search (e.g.\ \texttt{{[}1.0, 2.0, 3.0,
4.0, 5.0, 6.0{]}}), or an interval \texttt{\{"min": 1.0, "max":
10.0, "tol": 0.02\}} for continuous bounded optimization via Brent's
method. When power tuning is combined with top-$K$ pruning, the power
is optimized separately for each candidate value of $K_{\text{top}}$.

\paragraph{Top-$K$ pruning.}

The pruning level is set through \texttt{similarity\_top\_k}. It accepts
a fixed integer (e.g.\ \texttt{50}), a list of candidate values for
grid search (e.g.\ \texttt{{[}5, 10, 15, 20, 25, 30{]}}), or \texttt{null}
to disable pruning. The boolean \texttt{similarity\_top\_k\_symmetric}
determines whether the pruned graph is symmetrized, and \texttt{similarity\_top\_k\_chunk}
sets the number of rows processed per block during the top-$K$ selection.

\paragraph{Tuning procedure.}

When either \texttt{similarity\_top\_k} or \texttt{similarity\_power}
is given as a list, the implementation evaluates the full Cartesian
product of candidate values. If \texttt{similarity\_power} is given
as an interval, the continuous optimizer runs once per candidate value
of $K_{\text{top}}$. Each configuration is fitted and evaluated,
and the one that optimizes the metric specified by \texttt{selection\_metric}
is retained. The metric can be set to overall accuracy, Cohen's kappa,
the near-agreement rate, or the average negative log-likelihood. A
summary of every configuration tried is written to a CSV file for
post-hoc inspection.

\paragraph{Configuration summary.}

All options are controlled through a single \texttt{}~\\
\texttt{defaults.json} file. Table~\ref{tab:defaults_summary} lists
the keys relevant to the three strategies and the values each accepts.

\begin{table}[htbp]
\centering \caption{Key \texttt{defaults.json} options for the three tuning strategies.}
\label{tab:defaults_summary} {\small{}%
\begin{tabular}{V{\linewidth}l}
\hline 
{\small Option (accepted values) } & {\small Purpose }\tabularnewline
\hline 
\begin{cellvarwidth}[t]
{\small\texttt{similarity\_top\_k}}{\small{} }\\
{\small\emph{int, list, or }}{\small\texttt{\emph{null}}}{\small{} }
\end{cellvarwidth} & {\small Top-$K$ pruning level }\tabularnewline
\begin{cellvarwidth}[t]
{\small\texttt{similarity\_top\_k\_symmetric}}{\small{} }\\
{\small\emph{boolean}}{\small{} }
\end{cellvarwidth} & {\small Symmetrise the pruned graph }\tabularnewline
\begin{cellvarwidth}[t]
{\small\texttt{normalize\_similarities}}{\small{} }\\
{\small\emph{boolean}}{\small{} }
\end{cellvarwidth} & {\small Min--max rescale to $[0,1]$ }\tabularnewline
\begin{cellvarwidth}[t]
{\small\texttt{similarity\_power}}{\small{} }\\
{\small\emph{number, list, or dict}}{\small{} }
\end{cellvarwidth} & {\small Power $p$ (fixed, grid, or continuous) }\tabularnewline
\begin{cellvarwidth}[t]
{\small\texttt{embedding\_model}}{\small{} }\\
{\small\emph{string}}{\small{} }
\end{cellvarwidth} & {\small Any HF model; auto-detects ColBERT vs.\ embedding }\tabularnewline
\begin{cellvarwidth}[t]
{\small\texttt{colbert\_top\_k}}{\small{} }\\
{\small\emph{int or }}{\small\texttt{\emph{null}}}{\small{} }
\end{cellvarwidth} & {\small Candidate retrieval size in ColBERT mode }\tabularnewline
\begin{cellvarwidth}[t]
{\small\texttt{colbert\_batch\_size}}{\small{} }\\
{\small\emph{int}}{\small{} }
\end{cellvarwidth} & {\small Pairs per batch during MaxSim scoring }\tabularnewline
\begin{cellvarwidth}[t]
{\small\texttt{selection\_metric}}{\small{} }\\
{\small\emph{string}}{\small{} }
\end{cellvarwidth} & {\small Metric used to pick the best configuration }\tabularnewline
\hline 
\end{tabular}} 
\end{table}

\subsubsection{Weighted Estimation and Stability}

To address class imbalance in the rating categories, the script employs
balanced class weights. Each observation is assigned a weight inversely
proportional to the frequency of its category, so that the total contribution
of each category to the pseudo-likelihood is approximately equal.
This weighting improves the model's ability to predict minority categories,
at a possible cost in overall accuracy.

Numerical stability is ensured by clipping the predicted probabilities
to a small positive value (e.g., $10^{-15}$) before taking logarithms,
and by bounding the parameters to a reasonable range (e.g., $[-1000,1000]$).
These measures prevent underflow and divergence during optimization.

The implementation makes use of several Python libraries: NumPy (Harris
et al., 2020) for array operations, pandas (McKinney, 2010) for data
handling, and scikit-learn (Pedregosa et al., 2011) for evaluation
metrics. The script was developed using a ``vibe-coding'' approach,
in which a large language model (LLM) assisted in writing and refining
the code based on high-level descriptions of the statistical model
and desired functionality.

\section{Results Across Three Datasets}

\label{sec:results_three_datasets}

To evaluate the three similarity-sharpening strategies under a range
of conditions, we ran the full set of five variants on three datasets
that differ in size, response length, number of score categories,
and class balance.

The first dataset, referred to throughout as ILSA011, was assembled
in the context of a quality-control (QC) project for an international
large-scale assessment study. The responses were originally collected
in the source language and, for the present analysis, were machine
translated into English and subsequently screened to confirm that
the translated text contained no personally identifiable information
(PII). Only the translated and PII-checked (no cleaning was needed)
responses were used; no student identifiers or demographic information
were retained at any stage or used in any of the analyses. The score
variable used as the reference in all analyses reported here is the
human rater's score for each response; no model-generated or machine-assigned
score was used as the ground truth. The corpus contains $K=14{,}466$
responses scored on a three-level rubric.

The second and third datasets are drawn from the AERA (Automated Essay
scoring and Reasoning Assessment) corpus of Li and colleagues (Li
et al. 2023, Li 2024). We use Essays~5 and~6, each of which contains
student responses to a science or reasoning prompt scored on a four-point
rubric ranging from $0$ (lowest quality) to $3$ (highest quality).
Essay~5 comprises $N=1{,}238$ responses with a mean length of approximately
$143$ characters (SD~$=118$), while Essay~6 comprises $N=1{,}438$
responses with a mean length of approximately $141$ characters (SD~$=126$).
Both essay prompts exhibit substantial class imbalance, with the lowest
score category accounting for the majority of responses and the highest
category being comparatively rare. The two prompts also differ in
the consistency with which the scoring rubric appears to be applied,
which makes them a useful pair for evaluating how the model behaves
across differing degrees of rubric ambiguity. Table~\ref{tab:datasets}
summarizes the datasets.

\begin{table}[htbp]
\centering \caption{Datasets used in the comparison.}
\label{tab:datasets} %
\begin{tabular}{l|r|r|r|r}
\hline 
Dataset  & $K$  & Categories  & Mean length (SD)  & Imbalance\tabularnewline
\hline 
ILSA011 (GPT-4.1)  & 14{,}466  & 3  & 79.1 (39.9)  & moderate\tabularnewline
AERA Essay~6  & 1{,}438  & 4  & 141.4 (126.4)  & severe\tabularnewline
AERA Essay~5  & 1{,}238  & 4  & 143.0 (118.3)  & severe\tabularnewline
\hline 
\end{tabular}
\end{table}

\subsection{Result Summary}

Table~\ref{tab:master_results} presents the main results. The selection
metric in all analyses is Cohen's $\kappa$, so the variant chosen
by the tuning loop is the one that maximises chance corrected agreement
rather than the one that minimises cross-entropy. This will be discussed
some more in Section~\ref{sec:calibration_discrimination}.

\begin{table}[htbp]
\centering \caption{Main results for all five variants on the three datasets. Higher is
better for accuracy, kappa, near-agreement, and macro F1; lower is
better for the average negative log-likelihood. The best value in
each block is shown in bold.}
\label{tab:master_results} {\small{}%
\begin{tabular}{l|rrrrr}
\hline 
\multicolumn{6}{c}{{\small\textbf{ILSA011 ($K=14{,}466$, 3 categories)}}}\tabularnewline
\hline 
{\small Variant } & {\small Acc. } & {\small Kappa } & {\small Near } & {\small F1 } & {\small Loss }\tabularnewline
\hline 
{\small Full matrix (norm., $p=1$) } & {\small 0.815 } & {\small 0.700 } & {\small 0.986 } & {\small 0.739 } & {\small 0.504 }\tabularnewline
{\small Top-$K$ grid (sel.\ $K=34$) } & {\small\textbf{0.946}}{\small{} } & {\small\textbf{0.907}}{\small{} } & {\small 0.986 } & {\small\textbf{0.906}}{\small{} } & {\small 0.513 }\tabularnewline
{\small Power tuning ($p^{\ast}=8.53$) } & {\small 0.859 } & {\small 0.767 } & {\small 0.983 } & {\small 0.787 } & {\small\textbf{0.448}}{\small{} }\tabularnewline
{\small ColBERT (}{\small\texttt{mxbai}}{\small , fixed $K$) } & {\small 0.748 } & {\small 0.580 } & {\small 0.872 } & {\small 0.676 } & {\small 0.869 }\tabularnewline
{\small ColBERT (}{\small\texttt{mxbai}}{\small , $K$ grid) } & {\small 0.748 } & {\small 0.581 } & {\small 0.872 } & {\small 0.677 } & {\small 0.870 }\tabularnewline
\hline 
\multicolumn{6}{c}{{\small\textbf{AERA Essay~6 ($N=1{,}438$, 4 categories)}}}\tabularnewline
\hline 
{\small Variant } & {\small Acc. } & {\small Kappa } & {\small Near } & {\small F1 } & {\small Loss }\tabularnewline
\hline 
{\small Full matrix (norm., $p=1$) } & {\small 0.729 } & {\small 0.315 } & {\small 0.910 } & {\small 0.347 } & {\small 1.126 }\tabularnewline
{\small Top-$K$ grid (sel.\ $K=13$) } & {\small\textbf{0.837}}{\small{} } & {\small\textbf{0.496}}{\small{} } & {\small 0.939 } & {\small\textbf{0.511}}{\small{} } & {\small 1.132 }\tabularnewline
{\small Power tuning ($p^{\ast}=5.41$) } & {\small 0.784 } & {\small 0.388 } & {\small 0.933 } & {\small 0.428 } & {\small 1.043 }\tabularnewline
{\small ColBERT (}{\small\texttt{mxbai}}{\small , fixed $K$) } & {\small 0.821 } & {\small 0.464 } & {\small\textbf{0.944}}{\small{} } & {\small 0.499 } & {\small\textbf{0.985}}{\small{} }\tabularnewline
{\small ColBERT (}{\small\texttt{mxbai}}{\small , $K$ grid) } & {\small 0.821 } & {\small 0.463 } & {\small\textbf{0.944}}{\small{} } & {\small 0.498 } & {\small\textbf{0.985}}{\small{} }\tabularnewline
\hline 
\multicolumn{6}{c}{{\small\textbf{AERA Essay~5 ($N=1{,}238$, 4 categories)}}}\tabularnewline
\hline 
{\small Variant } & {\small Acc. } & {\small Kappa } & {\small Near } & {\small F1 } & {\small Loss }\tabularnewline
\hline 
{\small Full matrix (norm., $p=1$) } & {\small 0.493 } & {\small 0.125 } & {\small 0.822 } & {\small 0.253 } & {\small 1.233 }\tabularnewline
{\small Top-$K$ grid (sel.\ $K=34$) } & {\small\textbf{0.802}}{\small{} } & {\small\textbf{0.496}}{\small{} } & {\small 0.954 } & {\small\textbf{0.507}}{\small{} } & {\small 1.052 }\tabularnewline
{\small Power tuning ($p^{\ast}=7.90$) } & {\small 0.682 } & {\small 0.307 } & {\small 0.952 } & {\small 0.387 } & {\small 1.112 }\tabularnewline
{\small ColBERT (}{\small\texttt{mxbai}}{\small , fixed $K$) } & {\small 0.712 } & {\small 0.340 } & {\small\textbf{0.975}}{\small{} } & {\small 0.473 } & {\small\textbf{0.915}}{\small{} }\tabularnewline
{\small ColBERT (}{\small\texttt{mxbai}}{\small , $K$ grid) } & {\small 0.728 } & {\small 0.356 } & {\small 0.971 } & {\small 0.492 } & {\small 0.933 }\tabularnewline
\hline 
\end{tabular}} 
\end{table}

\subsection{Top-$K$ pruning provides highest Cohen's $\kappa$}

Across all three datasets, the top-$K$ variant achieves the highest
accuracy, Cohen's $\kappa$, and macro F1. The improvement over the
dense baseline that includes all similarities is clearly noticeable
in all three examples: For the ILSA011 case, the estimates accuracy
rises from $0.815$ to $0.946$ and kappa increases from $0.700$
to $0.907$; for the AERA Essay~6 example, accuracy rises from $0.729$
to $0.837$ and kappa from $0.315$ to $0.496$; for AERA Essay~5,
the accuracy estimate rises from $0.493$ to $0.802$ and kappa from
$0.125$ to $0.496$. In each case the improvement comes from a configuration
with a small number of neighbours per response---$34$ on ILSA011
and Essay~5, $13$ on Essay~6---rather than from a dense graph.

The magnitude of the gains is quite impressive on AERA Essay~5, where
the dense baseline essentially fails to separate the four score categories
(kappa $=0.125$) and top-$K$ pruning lifts it to a moderate agreement
($0.496$). This dataset was the hardest of the three: responses are
long and the score distribution is severely imbalanced. The result
suggests that when the semantic neighbourhood is noisy---as it tends
to be when responses are long and the classes are uneven---a sparse
local graph is not merely a computational saving but a substantive
modelling improvement.

\subsection{The optimal $K$ is task-dependent but modest}

The top-$K$ selection differs across datasets ($K=34$, $13$, and
$34$ respectively), but in all cases the chosen neighbourhood is
a very small fraction of the corpus. On ILSA011 with $K=34$, the
pruned graph has roughly $0.24\%$ of the edges of the full matrix.
On AERA Essay~6 with $K=13$, the figure is closer to $1\%$. These
numbers directly support the parsimony argument developed in the modelling
sections: a local Potts network with a few dozen neighbours per response
captures the scoring structure better than a dense graph in which
almost every edge carries negligible weight.

It is also worth noting that the optimal $K$ is not related to sample
size in our examples. The smallest dataset (Essay~5, $N=1{,}238$)
prefers a larger neighbourhood ($K=34$) than the medium-sized one
(Essay~6, $N=1{,}438$, $K=13$), which suggests that the relevant
scale is not the corpus size but the local semantic density of the
responses. in the case of short, tightly clustered, answers a smaller
neighbourhood suffices; in the case of long, heterogeneous answers
a larger neighbourhood is needed to capture the local structure.

\subsection{Power tuning is a consistent second place}

Continuous power tuning improves on the dense baseline on every dataset
while it never reaches the $\kappa$ or accuracy of top-$K$ pruning.
The selected exponents ($p^{\ast}=8.53$, $5.41$, and $7.90$) all
lie well above $1$, which indicates that in every case the model
benefits from sharpening the similarity distribution---emphasising
strong matches and suppressing weak ones. This is the same qualitative
effect that top-$K$ pruning achieves by a different mechanism (masking
rather than rescaling), and the two approaches should be regarded
as complementary rather than competing: power tuning adjusts the weights
on the edges that remain, while pruning retains only a small number
of the strongest edges.

An appealing practical property of power tuning is that it retains
the full graph, so it can be applied even when pruning would be undesirable
(for example, if the analyst wants to preserve a fully connected network
for interpretability of individual edges). Its consistent but smaller
gains over the baseline make it a reasonable first step when top-$K$
tuning is not available.

\subsection{ColBERT may help when responses are long}

The ColBERT variant, using the \texttt{mixedbread-ai/mxbai-edge-colbert-v0-32m}
checkpoint with a candidate pool of $K_{\text{ColBERT}}=100$ neighbours,
behaves quite differently across the three datasets. On the short
ILSA011 responses (mean $79$ characters) it is the weakest of the
four main variants (accuracy $0.748$, kappa $0.580$), clearly below
the dense baseline. On the two AERA essay prompts, however, ColBERT
is competitive: on Essay~6 it reaches accuracy $0.821$ and kappa
$0.464$, close to the top-$K$ winner; on Essay~5 it reaches accuracy
$0.712$ and kappa $0.340$, above the dense baseline and approaching
the top-$K$ winner.

This pattern is consistent with the mechanism of late interaction:
token-level alignment matters more when responses are long and information-rich,
and it matters less when responses are short and a single sentence
embedding already captures most of the semantic content. For short
constructed responses, we therefore recommend standard sentence embeddings
with top-$K$ pruning. For longer essays---especially when the scoring
rubric rewards fine-grained lexical or argumentative distinctions---ColBERT
is worth the additional computational cost.

A related finding is that the ColBERT $K$-grid variant barely improves
on the fixed $K_{\text{ColBERT}}=100$ configuration (difference in
$\kappa$ at the third decimal place on all three datasets). This
suggests that for the corpus sizes considered here, a fixed retrieval
pool of $100$ neighbours already captures the useful candidate set,
and further tuning of the ColBERT retrieval depth is not necessary.
The \emph{post-scoring} pruning controlled by \texttt{similarity\_top\_k},
by contrast, continues to matter: the ColBERT runs that also prune
the final graph to a small neighbourhood are the ones that perform
best on the essay datasets.

\subsection{Calibration and discrimination pull in different directions}

\label{sec:calibration_discrimination}

A consistent feature of the results is that the variant with the \emph{lowest}
average negative log-likelihood is not always the variant with the
\emph{highest} kappa or accuracy. On ILSA011, power tuning achieves
the lowest loss ($0.448$ versus $0.513$ for the top-$K$ winner)
while top-$K$ pruning achieves the highest kappa ($0.907$ versus
$0.767$). The same pattern appears on AERA Essay~6, where ColBERT
has the lowest loss ($0.985$) but top-$K$ has the highest kappa
($0.496$).

This divergence is expected and has a natural interpretation. Larger
neighbourhoods and denser similarity graphs produce smoother predicted
probabilities that track the empirical class distribution more closely,
which reduces the unweighted cross-entropy. But smoother probabilities
also compress the decision boundary, so fewer responses are assigned
to the minority classes---which is precisely where the agreement
metric is most sensitive. Sparser graphs do the opposite: they sharpen
the predicted probabilities and improve hard classification at the
cost of calibration.

For the purpose of reliability auditing, we regard \emph{discrimination
as the primary criterion} and \emph{calibration as a secondary one}:
the goal of the model is to recover the score assignment that a careful
rater would produce, not to reproduce the marginal distribution of
scores. This is why all runs in this comparison selected the best
configuration by maximising Cohen's kappa rather than by minimising
loss. Analysts who require well-calibrated probabilities for downstream
decisions should instead select by cross-entropy; the pipeline supports
both via the \texttt{selection\_metric} option.

\subsection{Summary}

Across three datasets that span short machine-generated answers and
long human-constructed essays, severe class imbalance and moderate
balance, three and four score categories, and corpora ranging from
about $1{,}200$ to over $14{,}000$ responses, the ranking of the
four strategies is remarkably stable: 
\begin{enumerate}
\item \textbf{Top-$K$ pruning} is the strongest option on every agreement
metric and every dataset, and the chosen neighbourhood is always a
small fraction of the corpus. 
\item \textbf{Power tuning} is a consistent second place, and its selected
exponent is uniformly well above $1$, indicating that sharpening
the similarity distribution is beneficial. 
\item \textbf{ColBERT} is competitive on long, information-rich responses
and adds little on short ones; a fixed retrieval pool of $100$ neighbours
appears sufficient. 
\item \textbf{The dense normalized baseline} is never the best option and,
on the hardest dataset, fails to separate the score categories at
all. 
\end{enumerate}
Taken together, these results make a clear empirical case for the
parsimony argument that motivates the Potts formulation: a sparse
local network of strongest semantic neighbours captures scoring reliability
better than a dense graph, and the improvement is largest precisely
where the dense graph struggles most.

\section{Conclusions}

This paper introduced the Potts auditing-reliability model (PARM),
a framework for multi-category scoring reliability that leverages
LLM-derived semantic similarities. The model defines a joint distribution
over scored responses using category-specific agreement indicators
and embedding-based pairwise weights. Its formulation naturally accommodates
any number of ordered or unordered rating categories, making it broadly
applicable to scoring tasks with multiple grade levels.

We estimated the model via pseudo-likelihood with balanced class weights
and analytical gradients, enabling efficient L-BFGS-B optimization.
Empirical evaluation on three datasets demonstrated its practical
utility and clarified the role of the similarity post-processing step.
On a corpus of $K=14{,}466$ machine-translated short answers scored
on a three-level rubric, top-$K$ pruning achieved an accuracy of
$0.946$ and a Cohen's kappa of $0.907$. On the two AERA essay prompts,
top-$K$ pruning achieved accuracies of $0.837$ (Essay~6, $\kappa=0.496$)
and $0.802$ (Essay~5, $\kappa=0.496$) on four-point rubrics with
substantial class imbalance. In every case the vast majority of misclassifications
occurred between adjacent score levels, confirming that the PARM captures
the ordinal structure of rubrics without imposing rigid equidistance
assumptions.

A central finding of this work is that the PARM is best understood
as a \emph{local} network model. Among the three strategies we compared
for sharpening the similarity signal---top-$K$ pruning, power transformation,
and ColBERT late interaction---top-$K$ pruning produced the highest
accuracy and Cohen's kappa on every dataset, and the neighbourhood
sizes selected by the tuning procedure were always a small fraction
of the corpus (between roughly $0.24\%$ and $1\%$ of the full pairwise
graph). This result makes an empirical case for the parsimony argument
that motivates the Potts formulation: a sparse network of strongest
semantic neighbours captures scoring reliability better than a dense
graph in which most edges carry negligible weight, and the improvement
is largest precisely on the datasets where the dense graph struggles
most.

Power tuning was a consistent second place and improved on the dense
baseline on every dataset. The selected exponents were uniformly well
above $1$, indicating that sharpening the distribution of similarities
is beneficial across a range of response types. Unlike top-$K$ pruning,
power tuning retains the full graph and therefore addresses the \emph{magnitude}
of the similarity signal rather than its \emph{structure}; the two
approaches are complementary and can be combined. ColBERT late-interaction
similarities were competitive on the longer AERA essays and added
little on the shorter ILSA011 responses, which is consistent with
the expectation that token-level alignment matters more when responses
are long and information-rich. A fixed candidate pool of $100$ neighbours
was sufficient for ColBERT in all cases, suggesting that further tuning
of the retrieval depth is not necessary at these corpus sizes.

The model parameters remain interpretable: $\beta_{c}$ reflects the
influence of agreement on category $c$, while $\gamma_{c}$ and $\mu_{c}$
capture total semantic similarity effects and category intercepts.
Across datasets, $\beta_{c}$ increased with higher categories, suggesting
that high-scoring responses exhibit stronger semantic coherence with
responses sharing the same score. A practical note that emerged from
the comparison is that the selection metric matters: cross-entropy
and Cohen's kappa can favour different configurations, because denser
graphs produce better-calibrated probabilities while sparser graphs
sharpen the decision boundary. For reliability auditing, where the
goal is to recover the score assignment that a careful rater would
produce, we recommend selecting the configuration by an agreement
metric such as $\kappa$ rather than by loss.

The PARM provides a flexible and extensible framework for reliability
auditing. Its formulation in terms of pairwise agreements and category-specific
weights readily generalizes to multiple raters by treating each rater's
score as a separate variable, and to hierarchical rating designs through
random effects for raters or items. Future work will explore these
extensions, along with applications to larger datasets and tasks requiring
fine-grained semantic distinctions.

\section*{References}

Byrd, R. H., Lu, P., Nocedal, J., \& Zhu, C. (1995). A limited memory
algorithm for bound constrained optimization. \emph{SIAM Journal on
Scientific Computing}, 16(5), 1190--1208.

Epskamp, S., Maris, G., Waldorp, L. J., \& Borsboom, D. (2018). Network
psychometrics. In P. Irwing, T. Booth, \& D. J. Hughes (Eds.), \emph{The
Wiley handbook of psychometric testing: A multidisciplinary reference
on survey, scale and test development} (pp. 953--986). Wiley Blackwell.
DOI:10.1002/9781118489772.ch30

Harris, C. R., Millman, K. J., van der Walt, S. J., et al. (2020).
Array programming with NumPy. \emph{Nature}, 585(7825), 357--362.

Hopfield, J. J. (1982). Neural networks and physical systems with
emergent collective computational abilities. Proceedings of the National
Academy of Sciences, 79(8), 2554--2558.

Ising, E. (1925). Beitrag zur Theorie des Ferromagnetismus. Zeitschrift
für Physik 31.1 (1925), pp. 253-- 258.

Khattab, O., \& Zaharia, M. (2020). ColBERT: Efficient and effective
passage search via contextualized late interaction over BERT. In Proceedings
of the 43rd International ACM SIGIR Conference on Research and Development
in Information Retrieval (pp. 39--48). Association for Computing
Machinery. https://doi.org/10.1145/3397271.3401075

Li, J., Gui, L., Zhou, Y., West, D., Aloisi, C., \& He, Y. (2023).
Distilling ChatGPT for explainable automated student answer assessment.
In Findings of the Association for Computational Linguistics: EMNLP
2023 (pp. 6007--6026). Association for Computational Linguistics.
https://doi.org/10.18653/v1/2023.findings-emnlp.399

Li, J. (2024). AERA: A dataset to enable LLMs for explainable student
answer scoring {[}Data set{]}. Hugging Face. https://huggingface.co/datasets/jiazhengli/AERA

McKinney, W. (2010). Data structures for statistical computing in
Python. \emph{Proceedings of the 9th Python in Science Conference},
445, 51--56.

Molenaar, P. C. M. (2004). A Manifesto on Psychology as Idiographic
Science: Bringing the Person Back Into Scientific Psychology, This
Time Forever. \emph{Measurement: Interdisciplinary Research and Perspectives,
2}(4), 201--218. DOI:10.1207/s15366359mea0204\_1

Rasch, G. (1960). Probabilistic models for some intelligence and attainment
tests. Danish Institute for Educational Research.

Reimers, N., \& Gurevych, I. (2019). Sentence-BERT: Sentence embeddings
using Siamese BERT-networks. \emph{arXiv preprint arXiv:1908.10084}.

Pedregosa, F., Varoquaux, G., Gramfort, A., et al. (2011). Scikit-learn:
Machine learning in Python. \emph{Journal of Machine Learning Research},
12, 2825--2830.

Potts, R.B., (1952). Some generalized order-disorder transformations.
Mathematical proceedings of the cambridge philosophical society. Vol.
48. 1. Cambridge University Press. 1952, pp. 106--109.

Razaee, Z. S., \& Amini, A. A. (2020). The Potts-Ising model for discrete
multivariate data. Advances in Neural Information Processing Systems,
33, 13727--13737. \\
 https://proceedings.neurips.cc/paper/2020/hash/9e5f64cde99af96fdca0e02a3d24faec-Abstract.html

Virtanen, P., Gommers, R., Oliphant, T. E., et al. (2020). SciPy 1.0:
fundamental algorithms for scientific computing in Python. \emph{Nature
Methods}, 17(3), 261--272.

von Davier, M. (2016). Rasch model. In W. J. van der Linden (Ed.),
Handbook of item response theory: Volume one: Models (pp. 31--48).
Chapman and Hall/CRC.

von Davier, M. (2018). Diagnosing Diagnostic Models: From von Neumann’s
Elephant to Model Equivalencies and Network Psychometrics. \emph{Measurement:
Interdisciplinary Research and Perspectives,} 16(1), 59--70. \\
 DOI:10.1080/15366367.2018.1436827

von Davier, M. (2026). Integrating network psychometrics and LLMs:
The Ising-Embeddings-Model applied to reliability auditing (Version
2). arXiv. https://doi.org/10.48550/arXiv.2608.26790 
\end{document}